# Transitions Between Cooperative and Crowding-Dominated Collective Motion in non-Jammed MDCK Monolayers

Steven J. Chisolm[1], Emily Guo[2], Vignesh Subramaniam[1], Kyle D. Schulze[2], Thomas E. Angelini[1,3,4]

[1]Department of Mechanical and Aerospace Engineering, University of Florida, Gainesville, FL, 32605.

[2]Department of Mechanical Engineering, Auburn University, Auburn, AL, 36849

[3]Department of Materials Science and Engineering, University of Florida, Gainesville, FL, 32605.

[4]J. Crayton Pruitt Family Department of Biomedical Engineering, University of Florida, Gainesville, FL, 32605.

**Transitions between solid-like and fluid-like states in living tissues have been found in steps of embryonic development and in stages of disease progression. Our current understanding of these transitions has been guided by experimental and theoretical investigations focused on how motion becomes arrested with increased mechanical coupling between cells, typically as a function of packing density or cell cohesiveness. However, cells actively respond to externally applied forces by contracting after a time delay, so it is possible that at some packing densities or levels of cell cohesiveness, mechanical coupling stimulates cell motion instead of suppressing it. Here we report our findings that at low densities and within multiple ranges of cell cohesiveness, cell migration speeds increase with these measures of mechanical coupling. Our observations run counter to our intuition that cell motion will be suppressed by increasingly packing or sticking cells together and may provide new insight into biological processes involving motion in dense cell populations.**

# 1. Introduction

The diversity of studies focusing on how cells move in condensed populations are often motivated by the long-standing recognition that collective cell motion plays a critical role in tissue development, health, and disease[1]. Early investigations of migration velocity fields within confluent cell islands uncovered connections to glassy-dynamics[2,3] and the jamming transition[4], in which motion in monolayers becomes arrested as cells pack more densely together. The reverse process has also been observed, in which jammed cells begin to move again following a large drop in exogenously applied hydrostatic pressure[5]. The intriguing connection between collective motion in condensed cell populations and arrested motion in other phases of inanimate condensed matter like molecular and granular materials motivated the development of numerous theoretical modeling approaches[6-8]. A few successful and widely adopted categories of theoretical models include the self-propelled particle models in which model cells are discrete objects[4,9], vertex models in which the model tissues are confluent tilings and the degrees of freedom are the vertices of the tiles[7,8,10], or self-propelled Voronoi models in which the degrees of freedom are the centers of confluent Voronoi tilings[6]. The convergence of theory and experiment has led to a growing understanding of solid-fluid transitions in tissues, though confluent tissues exhibit other collective behaviors that require further study, like giant density fluctuations[11] and mechanical coupling to large-scale intercellular fluid transport mediated by gap junctions[12-15].

While the study of motion in condensed cell populations has been motivated most often by its physiological importance, much of the early research in this area emerged from the field of single-cell mechanics, building on our understanding of the integrated relationship between the forces cells generate, the material properties of their environments, and the elasticity of cells themselves[16]. For example, simultaneous measurements of cytoskeletal elasticity and cell traction forces, complemented by studies of *in vitro* actin networks, showed that cell stiffness was controlled by motor-driven pre-stress in the cytoskeleton[17,18]. Completing this mechanical feedback loop, it was shown that the stiffness of the cell's substrate can modulate cytoskeletal pre-stress levels[19]. The experimental approaches and understanding developed in this area formed the foundation for careful studies of the traction forces exerted by cells within confluent cell islands on soft substrates, which uncovered how cells work together, each pulling on its neighbor and its substrate to maintain a state of collective tensile stress[20]. Providing insight into both single-cell and monolayer mechanics, observations of how cells interact dynamically with mechanical boundary conditions were made in single-cell stretching studies, where cell stiffening was again connected to intracellular pre-stress[21]. It was found that, in response to an imposed stretch, cells exhibit an increasing contractile force followed by a relaxation, with the whole cycle taking about 2 h. This active contractile response to externally applied stretches, controlled dynamically by the well-established mechanical behaviors of single cells, suggests that cells in monolayers may stimulate one another by a similar mechanism, pulling back and forth on one another as illustrated in Figure 1A. Likewise, groups of cells could cooperatively stimulate one another over larger scales through collective contraction as illustrated in Figure 1B,C; previous work showed that large neighboring patches of cells in monolayers cyclically expand and contract out of phase with one another[12]. Much like the crowding-dominated dynamics previously seen in monolayers as they approach a jammed state, we expect such cooperative collective dynamics to depend on cell density because it would arise from cell-cell mechanical coupling. Likewise, we expect cell cohesiveness to play an important role in cell-cell mechanical stimulation. However, collective motion driven by cooperative cell-cell mechanical interactions has not been previously observed, and a systematic exploration of the roles played by cell density and cell-cell cohesiveness in determining whether collective motion is cooperative or crowding-dominated has not yet been performed. The discovery of regimes of behavior in which cell migration speed grows with increasing cell packing density and cell cohesiveness would open the possibility to developing new interpretations of how collective motion in health and diseases arises in certain contexts and how to potentially control it.

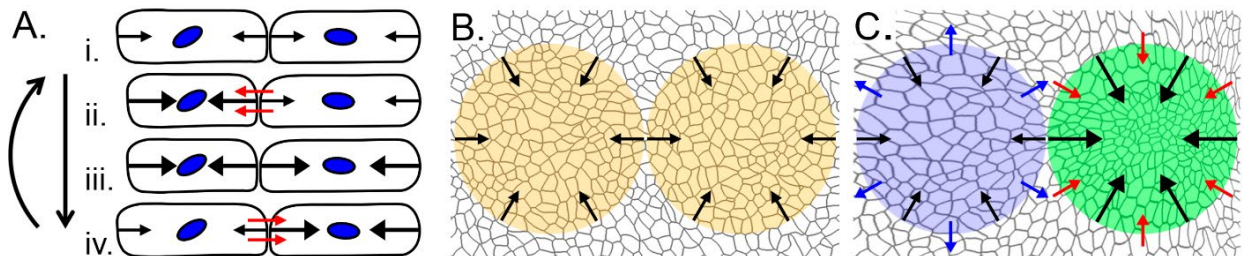

**Figure 1.** (A.) Cells contract in response to being stretched, suggesting that two neighboring cells in a monolayer could mechanically drive one another. Cells in mechanical equilibrium (*i.*) exhibit balanced contractile forces (black arrows). When forces become imbalanced (*ii.*), one cell contracts while the other is stretched (red arrows). Eventually, the stretched cell increases its contractile forces re-establishes balance (iii). Soon, the contracted cell relaxes while the stretched cell contracts (iv). The cycle then repeats. Applying this idea to patches of cells in a monolayer, two neighboring regions can start off in mechanical equilibrium (B.), but if one region expands (C, blue arrows) and the other contracts (C, red arrows), patches of cells may follow the cycle illustrated in (A.).

Here we systematically explore the conditions under which cell motion in monolayers appears to be suppressed by crowding or stimulated by cell-cell interactions. One of the control parameters we vary is cell density, and the two main variables we measure at different densities are the average cell migration speed and a parameter that quantifies cell shape, called the shape index. The other control parameter we vary is the level of cell-cell cohesiveness. By incubating monolayers in an E-cadherin antibody at many different concentrations we can reduce the levels of cell-cell cohesion[22,23]. We find, at all levels of cohesiveness, a very strong instantaneous anti-correlation between cell density and shape index. By contrast, we often find a lag between cell density fluctuations and changes in average migration speed. Mapping out the average cell migration speed in a 2D space of cell density and cohesiveness, we find multiple distinct regions where increased cell density and cohesiveness can either stimulate or suppress cell motion in all four different combinations. One general trend in this landscape is that the region of increasing migration speed with increasing cell density occurs at low densities; suppressed motion with increased crowding dominates at high densities as the monolayers likely approach jamming. We also find that at the highest levels of cohesiveness and the lowest levels of cohesiveness, migration speed increases with increasing cohesion; at intermediate levels of cohesiveness cell speed decreases with increasing cohesiveness. Thus, two different transitions separate these regimes of behavior. Examining the landscape of shape index in this 2D space, we find that the transition from stimulated to suppressed motion at low levels of cohesion correlates to strong changes in shape index, while the transition at high levels of cohesion is not accompanied by a strong change in shape index. Taken together, our results reveal that both packing density and cell cohesiveness can either promote or suppress motion in monolayers and that these transitions sometimes involve transitions in cell shape, but not always.

## 2. Results and Discussion

Previous work on jamming and glassy dynamics in cell monolayers demonstrated that crowding effects dominate cell motion at the higher end of cell densities. Thus, we expected that if a regime of motion dominated by cell-cell mechanical stimulation exists, it will be found at low cell densities, far below the jammed state, where the effects of crowding are reduced. To initiate experiments at the lowest possible confluent cell densities, we seed islands of Madin Darby canine kidney cells (MDCK) onto glass bottomed petri dishes at sub-confluent densities and allow them to fill in over the course of 12-24 h. Prior to cell seeding, the cells are coated with molecular collagen-1 to enable integrin-mediated attachment (see Methods). Once the monolayers have achieved confluence, the dishes are placed in a stage-top incubator on an inverted microscope and imaged in time-lapse for 24 to 48 h. We measure migration velocity fields using particle-image velocimetry (PIV) with PIVLab software, and we measure shape index and cell density using Cellpose image segmentation software[24]. The details of image processing and analysis can be found in sections 4.3 and 4.4 and the specific results subsections, below. As seen previously in MDCK monolayers[11-13], images of the monolayers exhibit large spatially varying density fluctuations that visibly correlate with large-scale patterns of motion (Fig 2). While previous work showed how cell density fluctuations couple to the velocity field, that work was done within a narrow window of average cell density; here we investigate how the overall magnitude of this motion depends on both cell density and cell-cell cohesion.

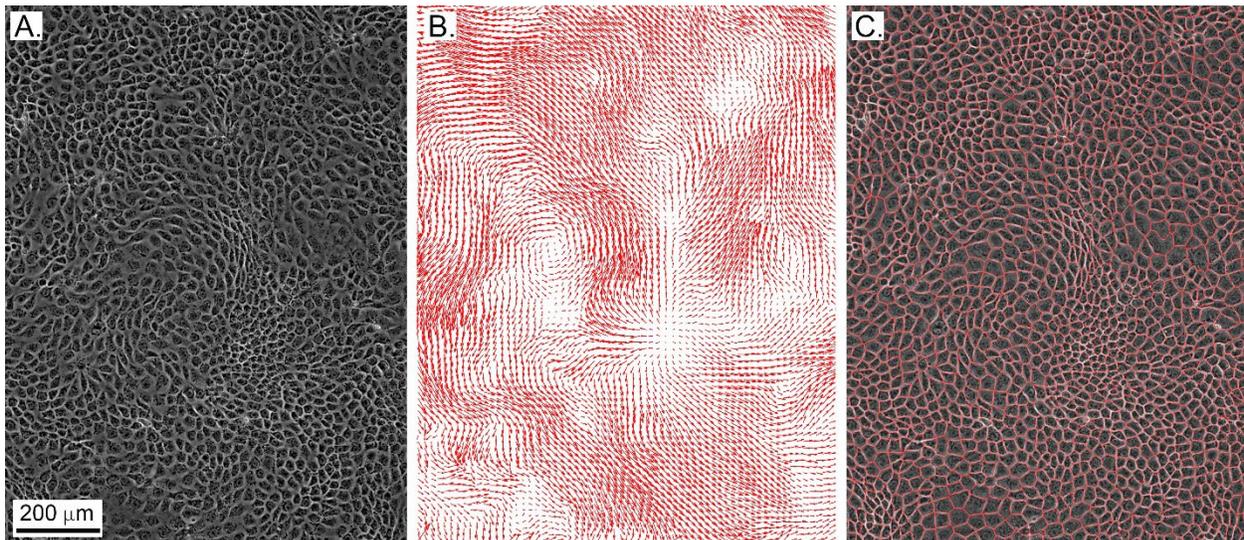

**Figure 2.** (A.) Giant density fluctuations are directly seen in phase-contrast images of MDCK monolayer islands, where cells in some regions are several times larger than in neighboring regions. (B.) Flow patterns in the migration velocity fields appear to spatially correlate to the density fluctuations, reflecting the known connection between cell density and motion. (C.) By segmenting phase-contrast images using Cellpose software, we measure the number of cells per unit area in the field of view and the shape index of each cell.

## 2.1 Correlations between cell density, migration speed, and shape index

To map out how cell motion, cell shape, and cell density vary over time and correlate with one another, we focus on spatially averaged parameters; our previous work focused on the spatial patterns. We choose the spatially averaged magnitude of the velocity field as our metric of average migration speed, given by $v(t) = \langle |\mathbf{v}(x,y,t)| \rangle_{x,y}$, where the absolute value is taken of each velocity vector, $\mathbf{v}$, and the angle brackets correspond to the mean computed over spatial locations, $x$ and $y$. We chose to take the mean after examining probability density functions of speed, finding a small amount of asymmetry but not enough to cause a significant difference between the mean and median (Fig. S1). For example, at the 5 h time-point in the data shown in Fig. 3A, the mean speed is 21.9 µm/h, the median speed is 19.7 µm/h, and the standard deviation is 12.7 µm/h (averaged over $9 \times 10^4$ velocity vectors). This standard deviation reflects the spatial variability in speed, not our confidence in the measurements; our control experiments consistently exhibit less than 5% RMS error in velocity measurements using PIVLab. Similarly, we choose spatially averaged shape index as our metric of cell shape, given by $q(t) = \langle q_j(t) \rangle_j = \langle p_j(t) / \sqrt{A_j(t)} \rangle_j$, where $p_j$ and $A_j$ are the perimeter and area of the $j^{\text{th}}$ cell, and the average is taken over all cells. Once again, to provide an example at the 5 h time-point, the probability density function of $q_j$ exhibits some asymmetry, the mean of $q_j$ is 3.87, the median $q_j$ is 3.81, and the standard deviation is 0.22 (averaged over 1238 cells). We determine the cell density by averaging over the cell-specific density, given by $\sigma(t) = \langle \sigma_j(t) \rangle_j = \langle 1/A_j(t) \rangle_j$. As with the other parameters at the same time-point, the probability density function of $\sigma_j$ is somewhat asymmetric, the mean of $\sigma_j$ is 1177 cells / mm$^2$, the median $\sigma_j$ is 1217 cells / mm$^2$, and the standard deviation is 417 cells / mm$^2$ (Fig. S1). Our control tests of the results from Cellpose software reveal an error of approximately 3% in $q$ and 5% in $\sigma$ (Fig. S2). Thus, we have a high degree of confidence in the time-traces shown in Figure 3.

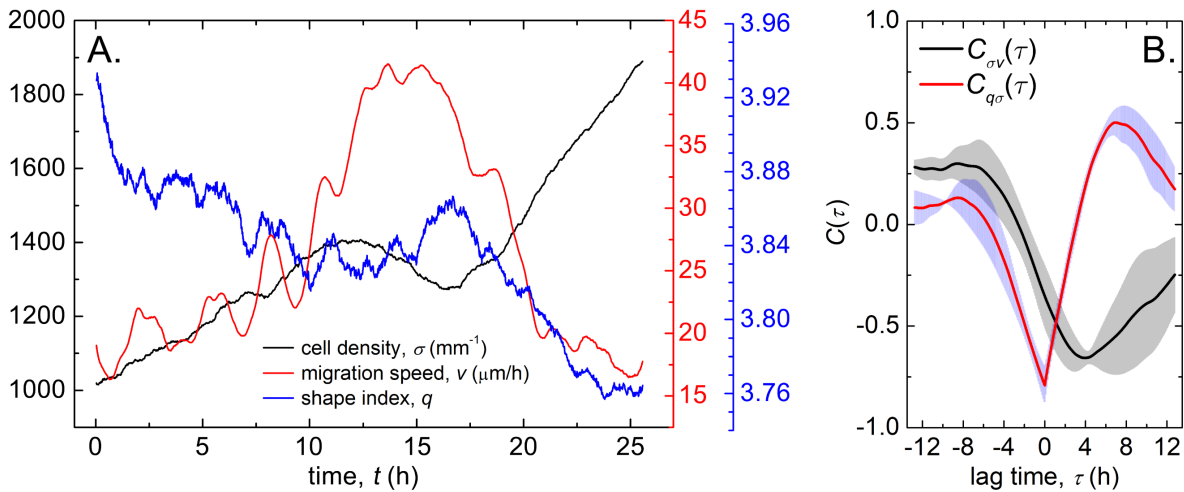

**Figure 3.** (A.) The three traces of spatially averaged parameters exhibit key trends such as decreasing $v(t)$ with increasing $\sigma(t)$ in the high-density range and increasing $v(t)$ with increasing $\sigma(t)$ in the low-density range. These two behaviors are consistent with crowding dominated motion at high densities and cell-cell stimulated motion at low densities. We also see that $q(t)$ and $\sigma(t)$ appear to be highly anticorrelated throughout the entire experiment. (B.) Cross-correlation functions of variable pairs show that indeed, $q(t)$ and $\sigma(t)$ are highly anticorrelated instantaneously, while $\sigma(t)$ and $v(t)$ do not exhibit a strong, clear, temporal relationship with one another, likely related to the two different regimes of behavior the cells exhibit in the two density ranges. (Errorbars represent the standard error about the mean, averaged across three separate replicate experiments.)

The three traces shown in Fig. 3 of $\sigma(t)$, $v(t)$, and $q(t)$ correspond to a monolayer with unmodified cohesiveness. For clarity, we do not display the standard deviation about the mean or any other metric of spatial variation at each time-point, given our findings about the distribution functions of these parameters. Examining $\sigma(t)$ and $v(t)$ simultaneously, we see the expected behavior at high densities, where speed monotonically decreases as density monotonically increases, and crowding dominates cell motion as the monolayer approaches a jammed state. By contrast, at low densities, we see both cell density and speed increasing with one another. This trend runs counter to crowding-dominated dynamics and may represent a regime of behavior where cells speed grows with increasing cell-cell stimulation. A large-scale fluctuation in cell density occurs in the time-period separating these two different categories of behavior. The detailed time-dependence of these parameters differs between monolayers, so to compare many different monolayers we eliminate time and investigate the direct relationships between $\sigma$, $v$, and $q$ in parametric plots, later in the manuscript.

Given that the shape index trace, $q(t)$, looks visually as if it would overlay the density trace, $\sigma(t)$, if flipped vertically, we chose to compute a cross-correlation function between the two variables. A normalized cross-correlation function between two variables, $A$ and $B$ for example, can be computed by first computing a normalized fluctuation of each variable about its mean, given by $\delta A(t) = \left(A(t) - \langle A(t) \rangle_t\right) / \sqrt{\langle A^2(t) \rangle_t - \langle A(t) \rangle_t^2}$, where we have subtracted the mean and divided by the standard deviation of $A$. The cross-correlation function is then given by $C_{AB}(\tau) = \langle \delta A(t) \delta B(t+\tau) \rangle_t$. Computed in this way, if at any time-shift, $\tau$, it is found that $C_{AB} = 1$, then $A$ and $B$ are perfectly correlated; if $C_{AB} = -1$, they are perfectly anti-correlated; if $C_{AB} = 0$ they are uncorrelated. Examining the cross-correlation function between shape index and cell density, $C_{q\sigma}(\tau)$, we see a strong negative peak at $\tau = 0$, where $C_{q\sigma}(0) = -0.79 \pm 0.09$ (mean ± standard error, averaged across three separate replicate experiments). This result shows that at any instant in time, fluctuations in $q$ and $\sigma$ are roughly 80% anti-correlated (Fig. 3B). We explore this striking relationship more in the next section by manipulating cell-cell cohesion. While a strong reciprocal relationship between shape index and density is expected from previous experimental[2] and theoretical[6] work on glassiness in monolayers, our result shows that even in the regime where motion is not dominated by crowding, the shape index and cell density are intimately linked and exhibit a similar reciprocal relationship. By contrast, from the density-speed correlation function, $C_{\sigma v}(\tau)$, we see that $\sigma$ and $v$ are far less anti-correlated instantaneously, where $C_{\sigma v}(0) = -0.35 \pm 0.15$ (averaged across three replicate experiments). We do see a moderately strong anti-correlation peak at $\tau = 3.95$ h equal to $-0.66 \pm 0.04$. In this case, it means that the $\sigma(t)$ trace is about 66% correlated with $-v(t + 3.95$ h) on average, or a relative increase in cell density leads to a corresponding drop in relative velocity about 4 h later. While we find this observation intriguing, in the investigations described next we could find no pattern in correlation strength or timing between $\sigma$ and $v$ when we varied cell-cell cohesiveness, but we did find the strong correlation between $q$ and $\sigma$ to be almost universal.

## 2.2 Dependence of speed and shape index on cell density and cohesiveness

In the vertex models of cell monolayer mechanics, cell-cell cohesiveness is one of the primary physical variables that control $q$, the shape factor; cells that are strongly cohesive are expected to exhibit higher $q$ values, increasing their perimeters relative to their areas. The strong anticorrelation we see between $q$ and $\sigma$ thus motivates us to test for how the trend may change when cell-cell cohesiveness is suppressed. We interfere with cell-cell cohesiveness by treating cells with an E-cadherin antibody that is known to block the formation of adherens junctions, DECMA-1[22,23]. We seed MDCK islands as described above and exchange their standard media with media containing DECMA-1, incubating for 2 hours before time-lapse imaging is commenced. Our experiments were performed at 10 different DECMA-1 concentrations between 0 and 10 µg/mL. While we are interested in the trends that may emerge across this range, we also perform three replicate experiments at DECMA-1 concentrations of 0, 5, and 7 µg/mL to assess variability at individual DECMA-1 concentrations. For each experiment we performed the same correlation analysis as described in section 2.1, again finding that $C_{q\sigma}$ exhibits a strong negative extremum at all DECMA-1 concentrations, with only one of the replicate experiments demonstrating a positive extremum (Fig. S3). Averaging across the 15 experiments that exhibit strong anti-correlations between $q$ and $\sigma$, we find the mean of the minimum correlation value of be $-0.88 \pm 0.14$ (mean ± standard deviation, $N = 15$). Thus, on average, $q$ and $\sigma$ are about 90% anticorrelated and exhibit relatively little variation across different conditions. By contrast, examining correlations between $\sigma(t)$ and $v(t)$, or between $q(t)$ and $v(t)$, we find the extrema of $C_{\sigma v}(\tau)$ and $C_{qv}(\tau)$ to be weaker and alternate between positively-correlated and anti-correlated behavior without any pattern or strong trend with varying DECMA-1 concentration. Thus, we focus on the strong anti-correlation between shape index and cell density, which can be directly seen in plots of $q$ versus $\sigma$ (Fig 4A). In the middle-range of cell densities, these $q$-$\sigma$ curves shift upward with increasing DECMA-1 concentration. We only show a few datasets for clarity; later we summarize all the datasets. Contrary to our assumption that increased cell cohesiveness drives an increase in cell perimeter, this trend indicates that *blocking* cadherins can result in increased perimeter for a given cell area (Fig 4a).

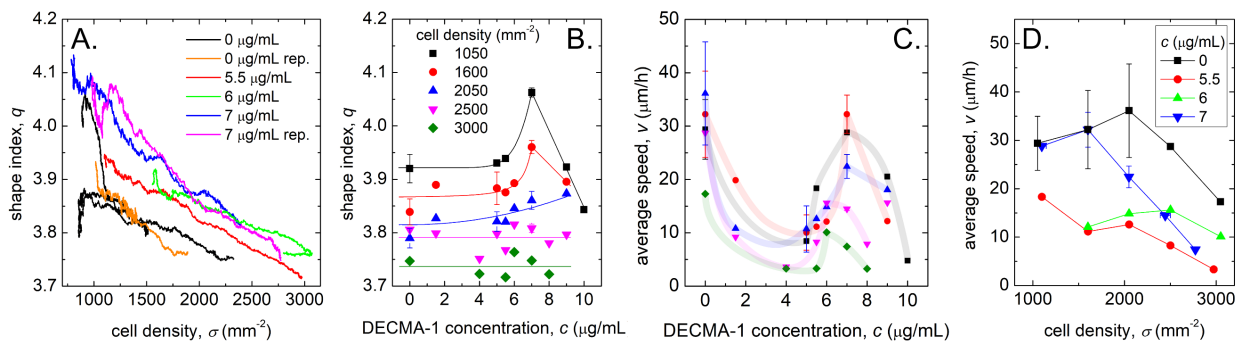

**Figure 4.** (A.) While $q(t)$ and $\sigma(t)$ can vary erratically in time, a parametric plot of $q$ versus $\sigma$ shows that the two variables are generally anti-correlated. In the middle-density range, the $q$-$\sigma$ curves shift upwards with increasing DECMA-1 concentration. (B.) The increase in $q$ with DECMA-1 concentration, $c$, is seen in the middle density range of $\sigma = 2050$ cells/mm$^2$. At higher densities we see no strong rise of $q$ with $c$; at lower densities we see a peak emerge near $c = 7$ μg/mL. (C.) At the same concentrations and cell densities as those shown in panel B (B and C share a legend), we see that cell speed first decreases with increasing $c$, then exhibits a local maximum near $c = 7$ μg/mL. (D.) Without manipulating cell cohesiveness ($c = 0$ μg/mL) we see the average cell migration speed rise then fall with increasing cell density (datapoints are mean ± standard deviation across three replicate experiments). Interfering with cell cohesion by adding DECMA-1 has a strong effect on the relationship between $v$ and $\sigma$; examining $v$ on the $\sigma$-$c$ landscape provides a clearer picture.

To systematically study how DECMA-1 interferes with cell cohesion and affects cell shape at different densities, we focus on a subset of datapoints from different experiments that can be grouped into single densities, $\sigma$, and plotted versus DECMA-1 concentration, $c$. Some monolayers never reached the higher densities while others could not be seeded confluently at low densities; the data displayed here represents the widest range of $\sigma$ we could achieve at each DECMA-1 concentration. We find that at the highest cell density where cells are approaching or may be in a jammed state, the shape index is the lowest and is relatively insensitive to DECMA-1 treatment, having an average value of 3.74 ± 0.02 (mean ± standard deviation), approaching the shape index of hexagonal packing, $q = 3.72$. At a slightly lower density we still see that the shape index is also relatively insensitive to DECMA-1 treatment, rising to an average value of 3.79 ± 0.02. We believe that cell crowding dominates in this high-density regime, pushing cells together toward hexagonal packing, regardless of their level of cohesiveness. In the middle range of densities, $\sigma$ = 2050 mm$^{-2}$, we see that $q$ rises with DECMA-1 concentration. This change in behavior occurs close to $q$ = 3.81, where a rigidity transition is predicted by the vertex model[10]. Finally, at the lowest densities we see that the $q$-$c$ curve continues to shift upward and develops a peak at 7 µg/mL. The rise in $q$ with both decreasing density and increasing DECMA-1 may be a consequence of reducing physical constraints on the cells, facilitating increased shape fluctuations. The emergent peak in the $q$-$c$ curves, however, shows that excessive reduction in cell cohesion reduces shape fluctuations, consistent with the possibility that cell cohesion facilitates cell-cell mechanical stimulation in this low-density regime (Fig 4b).

Since we found unexpected relationships between $q$, $\sigma$, and $c$, we chose to investigate how $v$ depends on the same combinations of $\sigma$ and $c$, now plotting $v$ instead of $q$. We expected $v$ and $q$ to follow similar trends, given the well-established connection between shape and motion in studies of jamming and glassy behaviors in monolayers. Indeed, we find that the $v$-$c$ curves peak near 7 µg/mL, comparable to how $q$ depends on $c$ and $\sigma$. Thus, in this regime of reduced cohesiveness, the faster moving cells are more irregularly shaped in general. By contrast, in the regime of higher cohesiveness between $c$ = 0 and 4 µg/mL, speed strongly decreases with increasing $c$, even though the $q$-$c$ curves are relatively flat in this regime; cells move faster with increased cell-cell cohesion without a dramatic evolution in shape (Fig 4c). The $v$-$\sigma$ curves also exhibit multiple regimes of behavior, depending on the concentration of DECMA-1. At $c$ = 0, where cell-cell cohesion is not manipulated, we see the datapoints reflecting the behavior observed in Fig 3a, where $v$ rises with density, reaches a peak, then drops as density rises further. Similarly, at $c$ = 7 µg/mL where the speed peaked in the low-density $v$-$c$ curves, we also see a peaked shape in the $v$-$\sigma$ curve. At intermediate DECMA-1 concentrations, we see that the $v$-$\sigma$ curves change only weakly or monotonically decrease with increased cell density (Fig. 4d). Together, these data indicate that the cell migration speed forms a hilly 2D landscape on the $c$-$\sigma$ plane, which we explore next.

## 2.3 Transitions between cooperative and crowding-dominated motion

Our analysis of how cell migration speed and shape index depend on both cell density and the concentration of added DECMA-1 reveals that multiple regimes of behavior exist where $v$ or $q$ can rise or fall with increasing $\sigma$ or $c$, depending on location in $\sigma$-$c$ space. A scatterplot of the $v$ datapoints in the $\sigma$-$c$ plane shows that the data form two hills centered around the densities and DECMA-1 concentrations where peaks were observed, above. To gain intuition about the shapes of these hills and to estimate the contours that separate different regimes of behavior, we fit a smooth 2D function through the data points. We find that the datapoints are well fit by the sum of two 2D Gaussian functions, each having different widths along their two orthogonal axes; we allow the centers, widths, and amplitudes of each Gaussian to vary freely as the error is minimized, resulting in $R^2 = 0.86$. Here, we are not interested in the fitting parameters themselves, but rather we seek a smooth 2D surface that captures the landscape the datapoints lay on (Fig 5A).

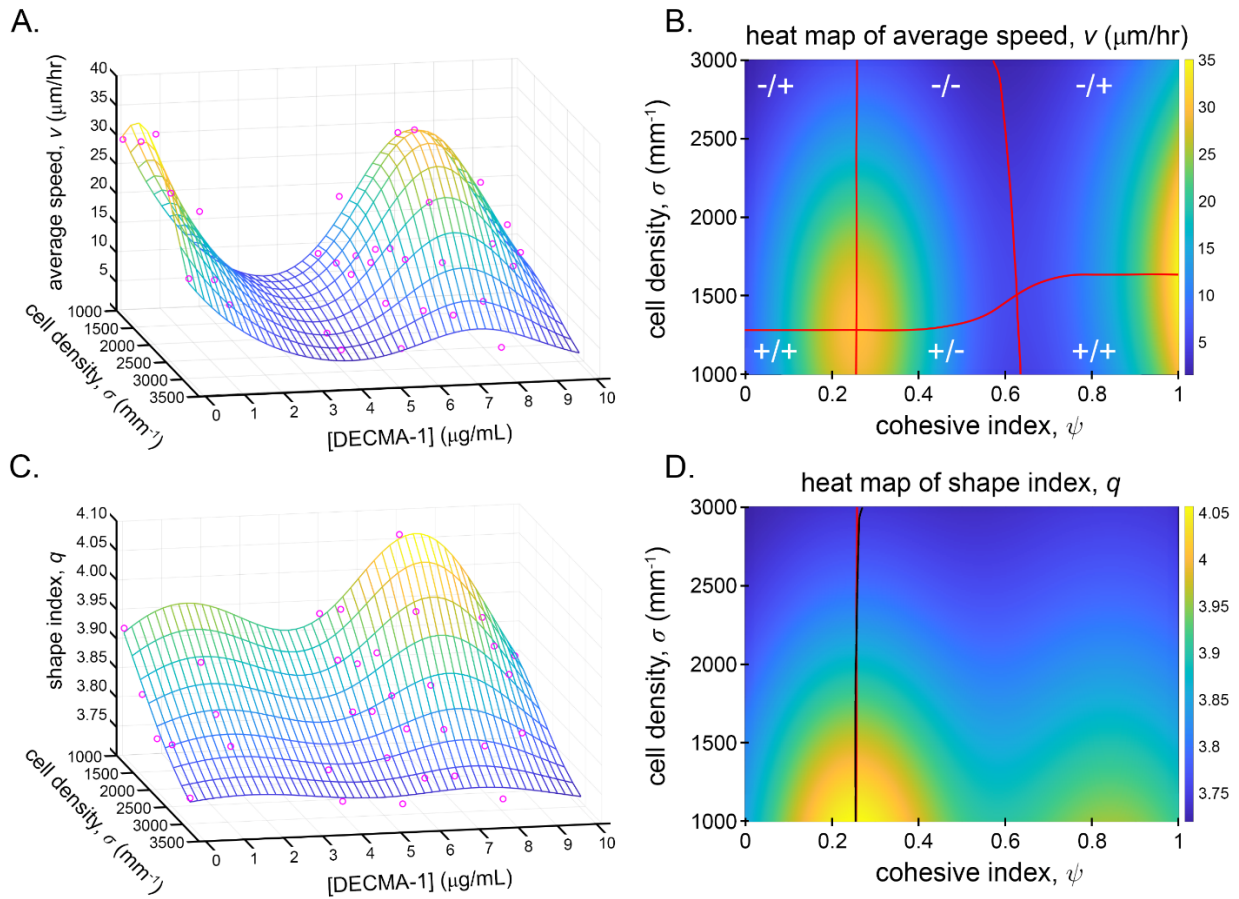

**Figure 5.** (A.) A scatterplot of cell migration speed, $v$, versus cell density, $\sigma$, and DECMA-1 concentration forms a hilly landscape. The contours correspond to the best fit function made from of the sum of two 2D Gaussians. (B.) A heat map of the best fit function from panel A, re-mapped onto the cohesive index, $\psi$. The red contours follow the ridges and valleys of the landscape. (C.) A scatterplot of cell shape index, $q$, versus $\sigma$ and DECMA-1 concentration also forms a hilly landscape, with reduced variation at low DECMA-1 concentrations. The contours correspond to the best fit function as used in panel A. (D.) A heat map of the best fit function from panel C, re-mapped onto $\psi$. The black contour follows the $q$-ridge that runs mostly along the $\sigma$ direction and the red contour is the corresponding ridge from panel B, showing that the $v$ border correlates to the $q$ border in this range of $\psi$. The "X/X" notations in panel B denote the different regimes of cell dynamics in the monolayer, where the first symbol denotes an increase or decrease in $v$ with $\sigma$ and the second symbol denotes an increase or decrease in $v$ with $\psi$.

To provide a clearer view of the migration speed landscape on the $\sigma$-$c$ plane, we create a heat-map from the best fit function (Fig. 5B). As part of this process, we map the $c$-axis to an adhesive index given by $\psi = (c_{max} - c)/c_{max}$, where $c_{max} = 10$ μg/mL. As a result, the maximal cohesive index of $\psi = 1$ corresponds to $c = 0$ μg/mL, and the minimal cohesive index of $\psi = 0$ corresponds $c = 10$ μg/mL. The two hills are easily seen in this representation, one with a ridge that runs along $\sigma$ at $\psi = 1$, which corresponds to untreated monolayers having maximal cohesiveness. Along this $\sigma$-oriented ridge we see the rise and fall in $v$, as seen in Figure 4D. The other hill in this landscape occurs at around $\psi = 0.27$, also rising and falling with increasing $\sigma$, close to the $c = 7$ μg/mL datapoints shown in Figure 4D. Examining this landscape along the $\psi$ direction, the same features are seen as those shown in Figure 4C at different densities. Thus, the smooth landscape created by our fitting procedure captures the complexity seen in the many datasets shown in Figure 4, but in a way that is easier to visually grasp.

To identify the boundaries between the different regimes of behavior separated by the hills on the $\psi$-$\sigma$ plane, we compute the gradient of $v(\psi,\sigma)$ and identify the contours along the ridges and valleys. Overlaying these contours on the heat map, we see that there are four different categories of behavior: (1) increasing $v$ along $\sigma$ and $\psi$, denoted as "+/+"; (2) increasing $v$ along $\sigma$ and decreasing $v$ along $\psi$, denoted as "+/–"; (3) decreasing $v$ along $\sigma$ and increasing $v$ along $\psi$, denoted as "–/+"; (4) decreasing $v$ along $\sigma$ and $\psi$, denoted as "–/–". We find that behaviors 3 and 4 both occur on the high-density side of the sigmoidal contour that is oriented along the $\psi$-axis; these behaviors all exhibit a decrease in cell motion with increasing density, as seen previously in studying glassiness and jamming in monolayers. By contrast, behaviors 1 and 2 both exhibit an increase in cell motion with increasing $\sigma$ and lay on the low-density side of the contour; these behaviors appear to reflect a regime where cells increasingly stimulate one another through increased interactions.

Examining the $v(\psi,\sigma)$ landscape along the $\psi$-axis, we again see regimes that appear to either suppress or promote motion with increased cohesiveness. At the lowest and highest ranges of $\psi$, cell migration speed increases with cell cohesiveness; in the middle range of $\psi$, cell migration speed decreases with cell cohesiveness. To gain insight into the mechanism that underlies the apparent reentrant behavior seen along the $\psi$-axis, we examine $q(\psi,\sigma)$, the shape index landscape in the $\psi$-$\sigma$ plane. We follow the same plotting and fitting procedure we performed in analyzing the $v(\psi,\sigma)$ landscape (Fig 5C,D). While, empirically, we see only one big hill in the landscape (Fig 4B), we fit the data to the same double-Gaussian function, finding a good fit having $R^2 = 0.95$.

The fitted surface indeed exhibits a much weaker bump at high cohesive index, so we focus our analysis on the larger peak in the lower range of $\psi$. Computing the gradient of $q(\psi,\sigma)$ in this lower range of $\psi$, we find no ridges or valleys running along the $\psi$-axis, but we identify a ridge that runs along the $\sigma$-axis that overlays almost perfectly with the corresponding ridge in $v(\psi,\sigma)$ space. The strong correlation between $q$ and $v$ along the $\psi$-axis in this range indicates that the physics of packing and geometry in monolayers may play a significant role in the transitions between different regimes of cell migration at moderate levels of cell-cell cohesion. By contrast, within the higher range of cohesive index, cell shape does not exhibit major changes with increasing $\psi$, yet $v$ grows dramatically. Thus, within the high-cohesiveness limit, we believe cell migration speed is enhanced dominantly by a mechanism of cell-cell stimulation. Finally, we note that the $q(\psi,\sigma)$ landscape seems to always slope downward with increasing cell density without any peaks along $\sigma$. This behavior is consistent with the strong, nearly universal, anti-correlation between $q$ and $\sigma$ we discussed earlier.

## 3. Conclusion

Here we have investigated how cell density and cell-cell cohesiveness influence cell motion in monolayers, seeking to identify regimes where these parameters serve to either promote or suppress cell motion. We were motivated by two major historical threads of thought in the broad field of cell mechanics. First, the foundational ideas about how cells sense and respond to the material properties of their surroundings[16] while actively and slowly responding to externally applied forces[21] led us to hypothesize that cells can stimulate one another in monolayers, potentially driving cooperative collective motion. Second, the extensively explored study of arrested motion in monolayers controlled by the physics of jamming, glassiness, packing, and geometry led us to ask how the trend toward arrested motion and jamming with increased packing density of cells could act antagonistically against motion promoted by cell-cell stimulation. By studying how the average cell migration speed and the average cell shape factor depend on both cell packing density and cell cohesiveness in monolayers, we have constructed a landscape that shows how cooperative interactions and crowding-dominated interactions create multiple regimes of cell monolayer dynamics, where different interactions appear to dominate.

Here we used the concentration of DECMA-1 as a metric of cell-cell cohesiveness. We recognize that blocking E-cadherin may produce indirect responses in the monolayer, and that it is preferable to connect DECMA-1 concentration to a quantitative level of cell-cell cohesion through direct measurement. As a possible alternative indirect metric, we measured the velocity-velocity correlation length for each experimental condition reported here, but found the average migration speed to exhibit highly erratic variation and no clear patterns when plotted against this correlation length. Thus, using DECMA-1 concentration as our cohesion metric, we find that confluent monolayer islands exhibit two regimes of motion in different ranges of cell density across all levels of cohesiveness. At low cell densities the migration speed, $v$, increases with increasing cell density, $\sigma$, while at high densities $v$ decreases with increasing $\sigma$. The crossover between these two classes of behavior depends on $\psi$, the level of cell-cell cohesiveness. Reducing $\psi$ shifts the boundary between these behaviors in the direction of lower $\sigma$. While this boundary is not the same as the boundary separating fluid from solid jammed phases found in previous work, we were curious whether this boundary occurs within a range of meaningful $q$ values. Focusing on the regime of $\psi$ where $q$ seems to play the biggest role, we identified the values of $v$ and $q$ that lay on this boundary up to $\psi = 0.65$. A plot of $v$ versus $q$ along this contour revealed a linear relationship that terminates exactly at the point $v = 0$ and $q = 3.81$ (Fig. S4). This is where a rigidity transition occurs in the vertex model[10], where unjamming occurred in experiments on lung epithelia[5], and where the glass transition boundary terminates in the self-propelled Voronoi (SPV) model[6] at $v = 0$. However, in contrast to the previous investigations, the boundary we have identified separates two classes of behavior that both occur in a fluid-like regime; arrested motion and jamming occurs at much higher cell packing densities where $q$ is much lower. For example, our boundary mapped into $v$-$q$ space lays in a $q$ range higher than 3.81 and slopes in the opposite direction from the glass transition boundary predicted by the SPV model. Our findings show that even in the fluid-like state in monolayers, there are multiple regimes of motion where migration speed can increase or decrease with cell density and cell cohesiveness.

Finally, across the cell densities explored here, we find that cell cohesiveness influences migration velocity in unexpected ways. We originally hypothesized that weakening cell-cell cohesion with the addition of DECMA-1 would inhibit the ability for cells to stimulate one another, possibly through out-of-phase cycles of contraction; we expected a monotonic decrease in $v$ with decreasing $\psi$. We find this trend to be the case in the high end of the $\psi$-axis, where decreasing cell cohesiveness from the untreated level reduces migration speed to a local minimum. This behavior does not correspond to a strong change in cell shape, which leads us to believe it is related to a reduction in cell-cell stimulation, comparable to the cooperative regime we see at low cell densities. While these behaviors are consistent with our hypothesis that cells in monolayers stimulate one another through mechanical forces, they deviate from current understanding and could arise from additional underlying mechanisms that we have not yet considered. Fortunately, current theoretical understanding and recent experimental results provide guidance for interpreting some parts of the $v(\psi,\sigma)$ landscape. For example, we did not anticipate a rise in migration speed followed by a second reduction as cell cohesiveness is further reduced. Since this trend is also seen in the shape index landscape within the same range of $\psi$, we expect that motion in this regime could be captured by existing theories in which $q$ is one of the dominating parameters that determines the collective dynamic state of a monolayer. We note that within this range of $\psi$, we see $q$ vary widely between approximately 3.7 and 4.1, and at the highest densities within this regime, $v$ becomes small, so the monolayer may exhibit signs of crossing into a jammed or otherwise arrested state. Indeed, our findings in this region of the $\sigma$-$\psi$ plane appear to align with previous work on airway epithelia where it was found that increased cell-cell "tugging" promoted unjamming, which correlates to an increased migration speed and shape index[5]. Outside this region of the $\sigma$-$\psi$ plane, our findings show that even within the fluid-like state of monolayer dynamics, there are multiple regimes of motion where migration speed can increase or decrease with cell density and cell cohesiveness. We believe these findings could help provide physical insight into transitions that occur in tissues associated with density change, such as the epithelial to mesenchymal transition (EMT) or the mesenchymal to epithelial transition (MET)[25]. While the EMT and MET have recently been thought of as transitions between fluid and solid states[26-28], our results show that a second transition could occur even within the fluid-like state and provide guidance on how to drive cells toward or away from a jammed state. The hilly landscapes of migration speed and shape index on the cohesiveness-density plane show that, counter to our intuition, sometimes increasing density or cohesiveness causes cells to move more rapidly rather than slowing them down.

## 4. Materials and Methods

### 4.1 Cell culture and cell island seeding

Madin Darby canine kidney (MDCK) epithelial cells are cultured in Dulbecco's modified Eagle's medium (DMEM) supplemented with 10% fetal bovine serum (FBS) and 1% penicillin streptomycin, maintained at 37 °C in a 5% $CO_2$ atmosphere. To prepare cell islands for experiments, the cells are grown to approximately 70% confluence, washed in phosphate buffered saline (PBS), and trypsinized. Once detached and dissociated, the trypsin is inactivated by 10-fold dilution with full media. The cells suspension is concentrated by gentle centrifugation and supernatant media exchange to achieve a chosen seeding density. A 50 μl drop of the solution is deposited near the center of a glass bottomed petri dish coated with molecular collagen-1 and incubated for 30 minutes to allow cell attachment. 2 mL of fresh media is added to the dish, which is then incubated for an additional 12-24 hours before commencing time-lapse imaging.

To prepare dishes coated with molecular collagen, 35 mm culture dishes having microscope coverslips as their base (Cellvis, product #:D35-20-1.5H) are exposed to 4.9 Watts UV light for 45 minutes. An acidic solution of molecular collagen at a concentration of 6 mg/mL ((Nutragen Type I Collagen Solution, Advanced Biomatrix 5010) is diluted with milli-Q water to a concentration of 0.38 mg/mL. 175 μL of the diluted collagen solution is pipetted onto the coverglass region of the dish and left at room temperature for 30 minutes. After incubation, the solution is removed and the dish is washed 2x with PBS buffer. The PBS buffer is removed and the dish is air dried before cell islands are deposited.

### 4.2 Microscopy

To perform time-lapse imaging on monolayer islands, the samples are placed in a stage-top incubator on an inverted Nikon Ti Eclipse microscope. The sample is maintained at 37 ºC in a 5% $CO_2$ atmosphere. Phase-contrast images are collected using a 10x objective at a magnification factor of 0.92 μm per pixel. Images are collected every minute for 24 to 48 hours. The samples are kept in focus using automated hardware.

### 4.3. Velocity field determination

We perform particle image velocimetry (PIV) to measure the cell migration velocity fields. Using PIVLab software, three levels of recursive fast Fourier transforms are computed on every sequential pair of images, yielding approximately 90,000 velocity vectors per frame with a spatial resolution of 4 pixels between vectors. We find that short timescale, sub-cellular motions add erratic fluctuations to the velocity vectors. Since we are interested in longer-timescale migratory motion, a 30-frame running boxcar average is performed on each vector to eliminate these short-time-scale fluctuations, as was done in previous work[2,12].

### 4.4 Image segmentation

We measure the shape and projected area of each identified cell by segmenting images of the monolayers islands using the Cellpose 2.0 software. Cellpose 2.0 uses a deep learning algorithm to perform image segmentation based on pre-trained or manually trained models[24]. We iteratively trained our own model to effectively segment phase contrast images of MDCK monolayers exhibiting a wide range of shapes and densities. Starting with the pretrained "cyto2_cp3" model, we segmented a small region of interest approximately 150 μm × 150 μm in size. After performing manual corrections to the segmentation, we use the original image and the segmented image to create a new model, trained over 100 epochs, that we then apply to a new image. We then use the two original images and the two segmented images to create and train another model. By repeating this process 32 times on an increasing diversity of images, iteratively building up a collection of validated segmented images and sequentially updated models, we arrive at a final model that is able to accurately segment all the images we collect in our experiments. We use custom written MATLAB code to determine the shape and area of each cell.

### 5. Acknowledgements

This material is based upon work supported by the National Science Foundation under Grant Number 2104429.

# Supplementary Information

**Transitions Between Cooperative and Crowding-Dominated Collective Motion in MDCK Monolayers**

Steven J. Chisolm[1], Emily Guo[2], Vignesh Subramaniam[1], Kyle D. Schulze[2], Thomas E. Angelini[1,3,4]

[1]Department of Mechanical and Aerospace Engineering, University of Florida, Gainesville, FL, 32605.

[2]Department of Mechanical Engineering, Auburn University, Auburn, AL, 36849

[3]Department of Materials Science and Engineering, University of Florida, Gainesville, FL, 32605.

[4]J. Crayton Pruitt Family Department of Biomedical Engineering, University of Florida, Gainesville, FL, 32605.

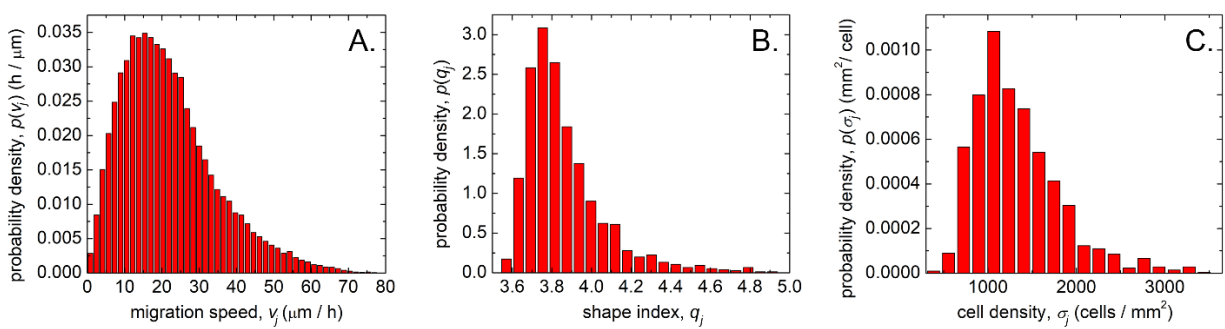

**Figure S1.** We examine how individual measurements of migration speed, $v_j$, shape index, $q_j$, and cell density, $\sigma_j$, are distributed by creating discrete probability density functions (PDFs) of the three variables. Each PDF is created by generating a histogram of constant bin-width and normalizing by both the bin-width and the total number of counts in the histogram. We find that all three PDFs are somewhat asymmetric, exhibiting tails at higher values of the independent variables. However, the asymmetry does not shift the mean value far from the median. (Panel A: $p(v_j)$ is generated from 89,873 speed measurements. Panels B,C: $p(q_j)$ and $p(\sigma_j)$ are generated from 1238 measurements of shape index and local density.

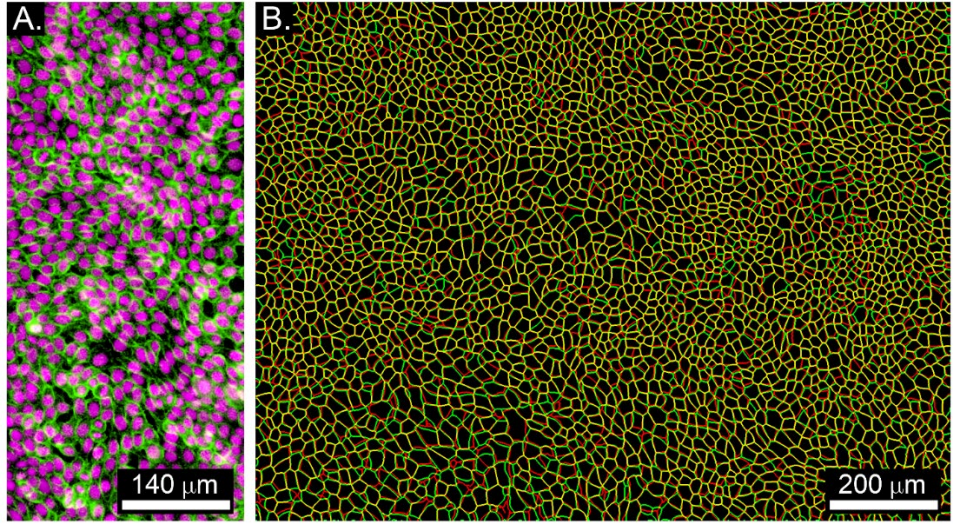

**Figure S2.** (A.) We check the segmentation accuracy of our custom-trained Cellpose 2.0 model by fixing and fluorescently staining monolayers and comparing results from fluorescence images to results from phase-contrast images. Segmentation maps using fluorescence images of nuclei (magenta) and E-cadherin (green) are made by employing pre-trained models. Segmentation maps using phase-contrast images are made using our custom-trained model. The details of our algorithm can be found in the Methods section of the main manuscript. (B.) Direct visual comparison shows that our custom-trained model for analyzing phase-contrast images produces cell-edges that overlay well with those produced from the fluorescence data and the pre-trained model (B). Yellow pixels correspond to the two segmentation maps overlaying; green pixels lay on the edges produced by phase-contrast analysis; red pixels lay on the edges produced by fluorescence image analysis. We find a 3% error in the average shape determined from phase-contrast images, relative to that determined from fluorescence images. We find a 5% error in the average cell density determined from phase-contrast images, relative to that determined from fluorescence images.

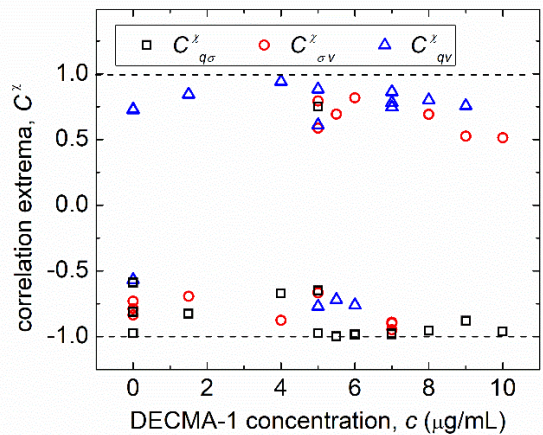

**Figure S3.** We examined the cross-correlation functions of all combinations of the three dynamic variables studied here: shape index, $q(t)$; cell density, $\sigma(t)$, and migration speed, $v(t)$. To summarize our findings, here we plot the values of the extrema found in the correlation functions, denoted by $C^\chi$. We find that in 15 out of 16 experiments, $C_{q\sigma}$ exhibits a strong negative peak, many of which lay close to a value of -1 (lower dashed line). We are not confident that that any pattern is found in the other cross-correlation functions.

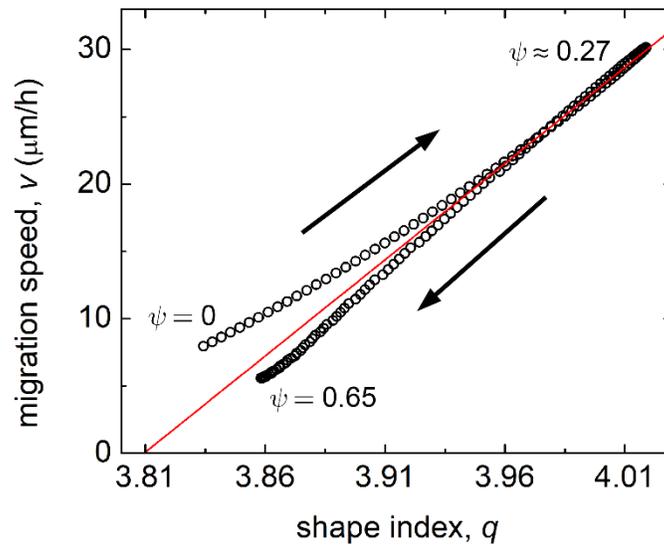

**Figure S4.** The landscape defined by $v(\psi,\sigma)$ exhibits a sigmoid-shaped boundary along the $\psi$-axis that separates a regime where $v$ increases with $\sigma$ from a regime where $v$ decreases with $\sigma$. To test whether this boundary occurs within a range of meaningful $q$ values, we extracted the corresponding points from the $q(\psi,\sigma)$ landscape and plotted the resulting $q$-$v$ contour. We truncated the contour above $\psi = 0.65$ since this is where a different regime of behavior emerges. Following the transition boundary from $\psi = 0$ to $\psi = 0.65$, we see $v$ and $q$ rise and fall with one another. A best fit line to this $q$-$v$ curve terminates at the point $v = 0$ and $q = 3.81$. While the monolayer in part of the $\psi$-$\sigma$ plane is in the fluid state, the termination of the extrapolating line at this point in $q$-$v$ space indicates to us that the system may be approaching the rigidity transition predicted by the vertex model at both ends of the contour.